\documentclass{phbauth}       
\usepackage{graphicx}
\usepackage{amsfonts}
\usepackage{latexsym}
\usepackage{amssymb}

\begin{document}
\begin{frontmatter}
\title{Quantum Hall effect today}
\author{V. J. Goldman\thanksref{thank1}}
\address{Department of Physics, SUNY, Stony Brook, New York 11794-3800, U.S.A.}
\thanks[thank1]{E-mail: goldman@insti.physics.sunysb.edu} 
\begin{abstract}
I present a brief survey of important recent developments in the quantum Hall effect. The review covers both fractional and integer regimes, from an experimentalist's perspective. The topics include direct measurements of fractional charge, composite fermion Fermi surface, spin textures, and edge state (chiral Luttinger liquid) dynamics.
\end{abstract}
\end{frontmatter}

{\bf 1. Introduction.} The last two decades since the discovery of the integer \cite {vonK} and fractional \cite {Tsui} quantum Hall effects (QHE) have witnessed remarkable increase in our understanding of low-dimensional strongly correlated systems. Many surprising and many beautiful experiments were performed (some of which are reviewed here), new theoretical understanding was gained,  new elegant and sophisticated techniques were developed, both experimental and theoretical.[3] 1985 and 1998 Nobel prizes in physics \cite {Nobel} were awarded to von Klitzing, and Tsui, Stormer and Laughlin for the discoveries.

The standard of resistance is now based on the integer QHE as $h/e^2 =25,812.807$ $\Omega$ {\em exactly}. The exactness of quantization of the Hall conductance is understood as a consequence of the gauge invariance of electromagnetic field and the quantization of the charge of current carriers.\cite {Lgauge,HalpEdge} Consider a gedanken experiment on a Corbino disc of two-dimensional (2D) electrons of areal density $n$ with quantizing magnetic field $B$ applied normal to the disc, so that Landau level filling is $\nu=hn/eB$. At a low temperature $T$ the system of electrons condenses into a QH state, integer or fractional, depending on $\nu$ and on how much disorder is present. First consider the case of the integer QHE, where the integer $i=1, 2, 3, ...$ is the number of occupied Landau levels; exact filling occurs at $\nu =i$. Now add adiabatically a flux quantum $\phi _0 =h/e$ in the inner hole of the disc; while $\phi _0$ is added one electron per occupied Landau level is transferred between the inner and the outer edges of the disc (provided they are connected by a wire). Since $\phi _0$ is added in the hole, the state of the electron system must be exactly the same as that before flux was added (gauge invariance). Thus, adding $\phi _0$ every $\delta t$, the Hall current is $ie/\delta t$, the voltage $\phi _0/\delta t$, and the Hall resistance $R _{xy} =\phi _0/ie$ (in 2D $\rho _{xy}= R _{xy}$). What is important here is that $R _{xy}$ remains quantized exactly even as $\nu$ is varied from the exact filling, because disorder localizes extra electrons or holes and the diagonal conductivity $\sigma _{xx}=0$.  

The fractional QHE at $\nu =f=\frac{i}{2pi+1}$ can be mapped onto the integer case using composite fermion (CF) theory.\cite {Laughlin,Jain,HLR} A composite fermion is an electron bound to an even number $2p$ ($p=1, 2, 3, ...$) of vortices of the many-particle wave function. The binding results from Coulomb interaction between the electrons, and it has been shown that the exact FQH ground states are very close to those in the CF theory, also the CF theory predicts the hierarchy of the FQH states observed in nature.\cite {JG} Since on the average, in an area, the number of vortices of the many-particle wave function is equal to the number of the flux quanta, in mean field theory CFs can often be thought of as electrons each binding $2p\phi _0$ of applied $B$. Thus CFs experience effective magnetic field $B _{cf}=B - 2pn\phi _0$ and the filling of the CF pseudo Landau levels $\nu _{cf}=n\phi _0/B_{cf}$ gives $\nu=\nu_{cf}/(2p\nu _{cf}+1)$. For $p=1$, for example, $B _{cf}=0$ occurs at $\nu =\frac{1}{2}$, thus the system of interacting electrons looks like zero-field metal of CFs. Also, the FQHE of interacting electrons at $\nu=\frac{i}{2i+1}$ maps onto the IQHE of non-interacting CFs at $\nu _{cf}=i$, with $i$ pseudo Landau levels occupied by CFs. 

\vspace{2 mm}
{\bf 2. Energy gaps.} The QHE itself, that is, the existence of plateaus in $R _{xy} = h/fe^2$ as a function of $\nu$, centered at exact filling $\nu =f$, depends critically on existence of a gap in the excitation spectrum. The physical nature of the gap can be distinguished between the single-particle gaps characteristic of the IQHE and the many-body interaction gap necessary for the FQHE. The examples of integer gaps are the kinetic energy gap between Landau levels, the spin flip gap, and the valley gap in Si. Experimentally, the fractional gap is always due to Coulomb interaction. It is quite common for the magnitude of the gap to be affected by several physical mechanisms, for example, spin flip gaps are often enhanced by the spin-dependent Coulomb exchange, and the fractional Coulomb gaps are reduced by mixing of several Landau levels.\cite{Books} When $\nu$ is varied from the exact filling quasiparticles are created (electron-like for $\nu >f$ and hole-like for $\nu <f$). The quasiparticles are localized by disorder and therefore no dissipative conduction is possible at $T=0$, $\sigma _{xx}=0$. At a finite $T$ quasiparticles are thermally excited in electron-hole pairs across the gap while maintaining overall charge neutrality. The thermally excited quasiparticles can carry dissipative current and, because $\sigma _{xx}$ is approximately proportional to their concentration, the activated transport experiments are used to determine the energy gap. The integer gaps were thus measured in a variety of samples in different materials, and a number of fractional gaps, in particular at $f=\frac{2}{3},\frac{3}{5},\frac{4}{7},\frac{5}{9},\frac{5}{11},\frac{4}{9},\frac{3}{7},\frac{2}{5}$ and $\frac{1}{3}$.\cite{Books,DuLeadley} The $\frac{1}{3}$ FQH gap was recently measured in inelastic light scattering experiments, a technique which also allowed to trace magnetoroton minimum in the excitation spectrum at finite wave vector.\cite{PinczukGap} 


\vspace{2 mm}
{\bf 3. Direct observation of fractional charge.} On a QH plateau quasiparticles are localized and their charge is well defined. In the case of the integer QH plateau with $\rho _{xy} = h/ie^2$ the quasielectrons are simply electrons in the $(i+1)^{st}$ Landau level, and the quasiholes are the holes (unoccupied states) in the $i^{th}$ level. In the case of the FQH plateau at $f =\frac{i}{2pi+1}$ quasielectrons are CFs (an electron binding $2p$ vortices) in the $(i+1)^{st}$ "Landau level" of composite fermions, and the quasiholes are the holes in the $i^{th}$ CF level. It has been predicted theoretically that the electric charge of these quasiparticles is $q=\frac{e}{2pi+1}$, positive for the quasiholes and negative for quasielectrons.\cite{Laughlin,Haldane,Halperin,GoldhJain} This charge fragmentation is a fundamental property of the FQH quantum fluid.

\begin{figure}[t]
\begin{center}
\includegraphics[width=0.75\linewidth]{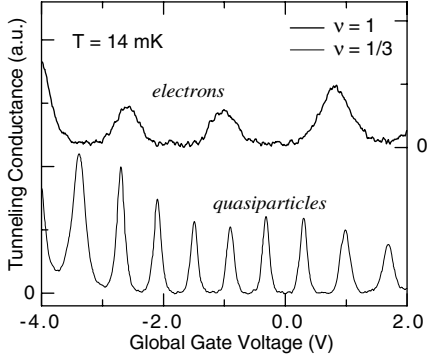}  
\caption{ The quantum antidot electrometer. Tunneling conductance peaks occur each time the occupation of the antidot changes by one particle: an electron for the IQH and a quasiparticle for the FQH regime. The charge of the particle is proportional to the global gate voltage; it takes same voltage to attract three quasiparticles as one electron.\protect \cite{SciCharge}}
\label{Fig. 1}
\end{center}
\end{figure}

The fractional charge of the quasiparticles has been measured directly in recent experiments.\cite{SciCharge} In the experiment, a 300 nm "quantum antidot" (potential hill) was etched into the layer of 2D electrons. In strong $B$ electron states circling the antidot are quantized by the Aharonov-Bohm condition that a state must contain an integer number of $\phi _0$. The occupation of the anitdot was detected via resonant tunneling: there is a peak in tunneling conductance each time a state becomes occupied, shown in Fig. 1. The surface area of the antidot $S$ was measured independently from the $B$-sweep data: the total flux through the antidot $BS$ changes by $\phi _0$ between two conductance peaks. There is a large global gate on the other side of the GaAs sample of thickness $d$; the gate forms a parallel plate capacitor with the 2D electrons. A gate voltage $V_G$ thus produces uniform electric field $E_{\perp} =V_G /d$, and induces a change of $\epsilon \epsilon_0 E_{\perp}$ in the surface charge density. The charge of one particle $q$ is thus directly given by the electric field needed to attract one more particle in the area $S$: $q=(\epsilon \epsilon_0 S/d)\Delta V_G$, where $\Delta V_G$ is the change of the global gate voltage between the successive conductance peaks. The results of the quantum antidot electrometer experiment were as follows: in the integer QH at $i=1$ and $2$ the charge of the electron was obtained as $1.57\cdot 10^{-19}$ Coulomb, the charge of the $f=\frac{1}{3}$ quasiparticles as $5.20\cdot 10^{-20}$ C, and of the $f =\frac{2}{5}$ quasiparticles as $3.1\cdot 10^{-20}$ C.\cite{SurfSciE} Subsequently, two groups reported determining the charge of $e/3$ quasiparticles using quite different technique: measuring shot noise as a function of current in constrictions defined in 2D electron layer.\cite{Noise}


\vspace{2 mm}
{\bf 4. Fermi surface of composite fermions.} The physics is different at even denominator fillings. As mentioned in Sect. 1 the effective field experienced by composite fermions $B_{cf}$ vanishes at $\nu =\frac{1}{2}$ and thus the system of interacting electrons looks like zero-field metal of CFs.\cite{HLR} The analogy is not exact, but still, the vortex attachment transformation somehow makes the many-body Coulomb interaction energy of 2D electrons look like kinetic energy of weakly interacting CFs, in a way not completely understood at present. However, various experiments give strong evidence for a reasonably well defined Fermi surface of CFs of Fermi wave vector $k^{cf}_F$ near $\nu =\frac{1}{2}$. If $B_{cf}=0$, CFs experience no external magnetic field and move in straight lines. At small $B_{cf}$, CFs execute cyclotron motion on the Fermi surface. The experiments detect, by different techniques, the resonance occuring when the CF cyclotron radius $R^{cf}=\hbar k^{cf}_F /eB_{cf}$ is commensurate with an external length $L$. Note that for 2D spin-polarized particles $k^{cf}_F=\sqrt{4\pi n}$ depends only on density.

\begin{figure}[b]
\begin{center}
\includegraphics[width=0.68 \linewidth]{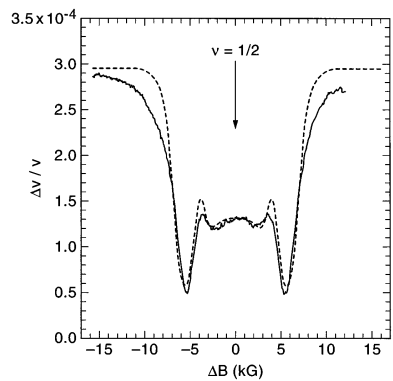} 
\end{center}
\caption{Relative change of propagation velocity {\em vs.} CF effective magnetic field for 8.5 GHz SAW. The dashed line is a model fit including inhomogeneous broadening.\protect \cite{Willett95}}
\label{Fig. 2}
\end{figure}

In surface acoustic wave (SAW) experiments the velocity of propagation and the attenuation of SAW is affected by their interaction with 2D electrons. Anomalies in propagation of high frequency SAW, in disagreement with what was expected from $dc$ conductivity, were observed near $\nu =\frac{1}{2}$.\cite{Willett90} These anomalies were successfully explained in terms of interaction of small wave vector SAW with a gapless Fermi sea of CFs, at a mean field level.\cite{HLR} In subsequent experiments a geometric resonance has been observed, Fig. 2, when the CF cyclotron diameter is equal to the SAW wavelength.\cite{Willett95} In effect, the SAW sets up a lateral superlattice potential, and similar commensurability geometric resonances were observed in experiments with lateral superlattices defined by lithography, as well as in samples with lithographic 2D arrays.\cite{KangSmet}

\begin{figure}[t]
\begin{center}
\includegraphics[width=0.87\linewidth]{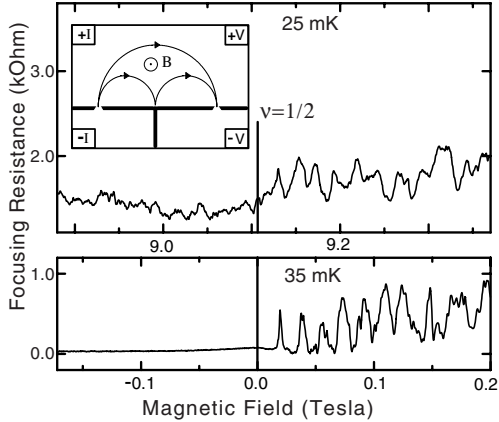} 
\end{center}
\caption{ Magnetic focusing of composite fermions near $\nu =\frac{1}{2}$ compared with focusing of electrons near $B=0$. Composite fermions experience $B_{cf}=B - B(\nu =\frac{1}{2})$. The two $B$ axes differ by $\sqrt 2$ to account for spin polarization at high $B$. Inset shows the focusing sample geometry, where two possible focusing paths are shown by arrows.\protect \cite{CFfocus}}
\label{Fig. 3}
\end{figure}

Another type of experiment where CFs were detected is magnetic focusing.\cite{CFfocus} Magnetic focusing of 2D electrons has been studied extensively.\cite{vanHouten88} As shown in Fig. 3, a current is passed through the left (emitter) constriction and the voltage across the right (collector) constriction is measured. In the linear regime, the voltage is proportional to the applied current, and  the ratio is defined as nonlocal focusing resistance. Classically, current carriers emitted out of the left constriction execute cyclotron motion on the Fermi surface, and for most angles of injection are focused onto the collector when $B$ is such that the cyclotron diameter $2R$ is equal to the constriction separation $L$. Allowing for $j-1$ specular reflections, a series of focusing peaks occurs when $2jR=2j\hbar k_F /eB=L$. As a function of $B$ the peaks are nearly periodic, $B_j =j\Delta B$, spaced by $\Delta B=2\hbar k_F /eL$, Fig. 3, both for electrons near $B=0$ and for CFs near $\nu =\frac{1}{2}$. In the opposite direction of $B$ the current carriers are deflected to the left, and no focusing is observed.


\vspace{2 mm}
{\bf 5. Edge state transport.} The edge of a QH system plays a central role in charge transport because edge confining potential prevents localization of electron states by disorder, and applied current is carried in a sample by the delocalized 1D edge states.\cite{HalpEdge,Books} Dissipationless conduction by edge states has been established in experimental observation of dramatic "nonlocal resistance" in macroscopic QH samples.\cite{McEuen,JKWang} The direction of circulation of edge excitations is determined by the applied magnetic field, and the theoretical picture of QH edge is based on chiral Luttinger liquid ($\chi $LL) models.\cite{Wen,Books} $\chi$LL theories make several dramatic predictions for edges of the FQHE at $\nu =f$. For example, for $f=\frac{1}{3}$ the width of a resonant tunneling peak should scale as $\propto T^{\frac{2}{3}}$ in the low voltage Ohmic regime $eV<2\pi k_BT$, in contrast to the familiar Fermi liquid $T^1$ scaling. The range of applicability of the $\chi $LL behavior is not well known however; a $\propto T^1$ dependence was reported in experiments on $e/3$ quasiparticle resonant tunneling.\cite{Maasilta}

\begin{figure}[t]
\begin{center}
\includegraphics[width=0.76\linewidth]{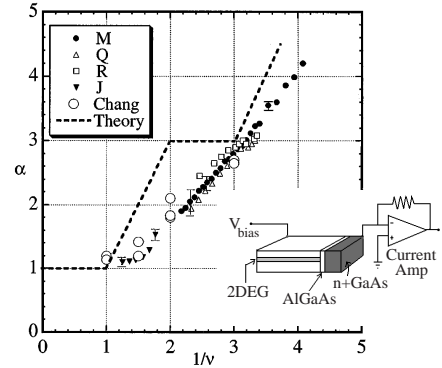} 
\end{center}
\caption{The tunneling $I\propto V^{\alpha}$ power law exponent {\em vs.} the inverse filling $1/\nu $ for five $\bullet \triangle {\tiny \Box \bigtriangledown }\circ $ samples. The dashed line gives the $\chi $LL-theoretical prediction. A schematic of the cleaved-edge overgrown sample and the $I-V$ measuring circuit are shown above.\protect \cite{Chang98}}
\label{Fig. 4}
\end{figure}

Another dramatic prediction of $\chi $LL theories is that for electron tunneling into a FQHE at $\nu =f$, at high bias $eV>2\pi k_BT$ the tunneling current is non-Ohmic, with a power law dependence $I\propto V^{\alpha}$. For $f =\frac{i}{2pi \pm1}$ the exponent is predicted to be $\alpha=2p+1$ for plus, and $\alpha=2p+1-\frac{2}{i}$ for minus in the denominator, shown in Fig. 4. The experiment was performed on a cleaved edge 2D heterostructure overgrown by a metallic $n+$ GaAs on top of an AlGaAs barrier.\cite{Chang98} $I-V$ measurements found power law tunneling, however with $\alpha$ a continuously increasing function of $1/\nu$. This behavior is not understood at present, in particular, why $\alpha $ varies on a QH plateau and what plays role of edge states in gapless regions, as near $\nu =\frac{1}{2}$.


\vspace{2 mm}
{\bf 6. Spin textures.} In the picture of independent spins, each Landau level is split into spin up and spin down levels, so that electrons are fully spin polarized for $\nu <1$, completely unpolarized at even filling and partially polarized otherwise. Electron-electron interaction, however, is spin-dependent, which leads to spin-dependent correlations of the many electron states. This effect is particularly strong in GaAs, where the band $g$-factor is small, 0.44. The resulting spin depolarization depends self consistently on $\nu$, which affects the Coulomb correlations, as was studied in recent photoluminescence experiments.\cite{Kukushkin} Since spin Zeeman energy responds to total magnetic field, while the filling $\nu$ of 2D electrons is determined only by the normal component of $B$, certain QH states undergo phase transitions from un- or partially-spin polarized to fully polarized in tilted $B$.\cite{ClarkEisensteinDu} 

\begin{figure}[t]
\begin{center}
\leavevmode
\includegraphics[width=0.70\linewidth]{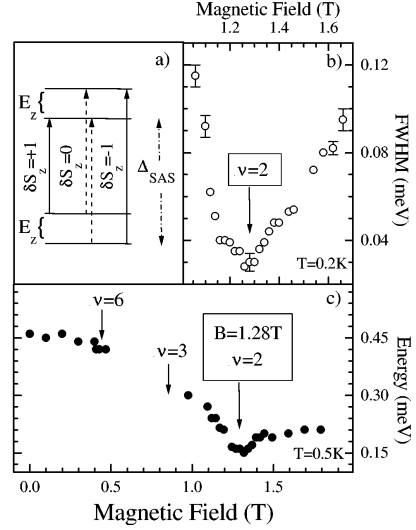} 
\caption{The width and the energy of the spin-density excitation (SDE) in a double layer sample. SDE is a linear combination of $\delta S_z=0$ transitions, dashed lines; the spin flip excitations are $|\delta S_z|=1$. $E_z$ is the Zeeman splitting, and $\Delta_{SAS}$ is the symmetric-antisymmetric gap.\protect \cite{Pinczuk97}}
\label{Fig. 5}
\end{center}
\end{figure}

Evidence for even more exotic topological spin textures \cite{Books} has been reported in recent experiments. In double  electron layer samples interlayer spin-dependent correlations can be nearly as strong as intra layer correlations, even when tunneling between layers is small. A remarkable softening of long wavelength intersubband spin excitations, Fig. 5, was observed at $\nu =2$.\cite{Pinczuk97} This occurs when each layer has only one spin-split Landau level occupied. Probed by resonant inelastic light scattering, these spin density wave excitations soften to as much as 0.1 of the $B=0$ values. Even though further studies are required, these observations indicate magnetoroton instability precursor to a phase transition to an ordered spin phase, such as quantum antiferromagnets, long predicted for coupled QHE systems.

\begin{figure}[b]
\begin{center}
\leavevmode
\includegraphics[width=0.73\linewidth]{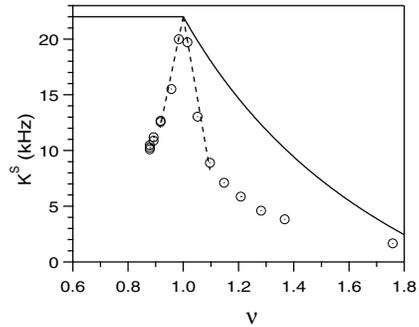} 
\caption{Electron spin polarization {\tiny $\odot $} measured by NMR Knight shift near $\nu=1$ at $B=7$ T, $T=1.5$ K. The solid line assumes non-interacting electrons, the dashed line is a fit for $S=3.6$ finite size skyrmions.\cite{Barrett95}}
\label{Fig. 6}
\end{center}
\end{figure}

In single electron layers at $\nu \leq 1$ electrons are fully spin polarized for large $g$-factors or in the limit $B \rightarrow \infty$. For small $g$-factors topological charged defects in spin orientation called skyrmions are possible.\cite{Sondhi,Books}  To visualize a skyrmion one can think of all far spins pointing up, then rotate smoothly downward at the position of the skyrmion. The size of a skyrmion is controlled by the competition between the Zeeman energy, which tends to reduce the number of flipped spins, and the Coulomb exchange energy, which tends to spread the extra charge over a large area. In $\nu =f$ QHE, skyrmion excitations are predicted to have net charge of $fe$, in general  different from charge of a Laughlin quasiparticle. Evidence for skyrmions was observed in NMR experiments measuring Knight shift of $^{71}$Ga nuclei, Fig. 6, which is sensitive to the spin polarization of 2D electrons. The spin polarization exhibits a maximum at $\nu =1$, and falls off steeply on either side. This behavior is understood as due to a finite size ($S\sim 4$) skyrmions excited as $\nu $ is varied from the exact filling. More recently, evidence for formation of collective spin textures, though even of smaller size, was obtained for FQHE, $\nu =\frac{1}{3}$.\protect \cite{Barrett95}


\vspace{2 mm}
{\bf Acknowledgments: } The author benefited greatly from many discussions with B. I. Halperin, S. M. Girvin, N. Read, J. K. Jain, D. C. Tsui, A. H. MacDonald, S. A. Kivelson, H. L. Stormer, I. L. Aleiner, X. G. Wen and many other colleagues. This work was supported in part by the NSF under Grant No. DMR-9629851. 


\end{document}